\RequirePackage{lineno}
\documentclass[aps,twocolumn,superscriptaddress,showpacs,amsmath,amssymb,nofootinbib]{revtex4}
 \pdfoutput=1
 \usepackage{graphicx,epsf,amsmath}
 \usepackage[]{latexsym}
 \usepackage{amsmath}
 \usepackage{subcaption}
 \usepackage{float}
 \usepackage{color}
 
 \usepackage{dcolumn}   
 \usepackage{bm}        
 \usepackage{amssymb}   
 \usepackage[utf8]{inputenc}

 \newcommand{\be}{\begin{eqnarray}}
 \newcommand{\ee}{\end{eqnarray}}

\begin{document}
 
 
\title{Restoration of azimuthal symmetry of muon densities in extended air showers}

 \author{Nicusor Arsene}
 \email[]{nicusorarsene@spacescience.ro}
 \affiliation{Institute of Space Science, P.O.Box MG-23, Ro 077125 
 Bucharest-Magurele, Romania}

 \author{Markus Roth}
 \email[]{markus.roth@kit.edu}
 \affiliation{Karlsruhe Institute of Technology, Institut für Kernphysik, Karlsruhe, Germany}

 \author{Octavian Sima}
 \email[]{octavian.sima@partner.kit.edu}
 \affiliation{Physics Department, University of Bucharest, 
 Bucharest-Magurele, Romania}
 \affiliation{“Horia Hulubei” National Institute for Physics and Nuclear Engineering, Romania}
 \affiliation{Extreme Light Infrastructure - Nuclear Physics, ELI-NP, Ro 07725 Bucharest-Magurele, Romania}

 \date{\today}

\begin{abstract}
At ground level, the azimuthal distribution of muons in inclined Extensive Air Showers (EAS) is asymmetric, mainly due to geometric effects.
Several EAS observables sensitive to the primary particle mass, are constructed after mapping the density of secondary particles from the 
ground plane to the shower plane (perpendicular to the shower axis). A simple orthogonal projection of the muon coordinates onto this plane distorts 
the azimuthal symmetry in the shower plane. Using CORSIKA simulations, we correct for this distortion by 
projecting each muon onto the normal plane following its incoming direction, taking also into account the attenuation probability.
We show that besides restoring the azimuthal symmetry of muons density around the shower axis, the application of this procedure  has a significant impact on the reconstruction of the distribution of the muon production depth and of its maximum,
$X_{\rm max}^{\mu}$, which is an EAS observable sensitive to the primary particle mass. Our results qualitatively suggest that not including it in the 
reconstruction process of $X_{\rm max}^{\mu}$ may introduce a bias in the results obtained by analyzing the actual data on the basis of Monte Carlo 
simulations. 

\end{abstract}


\maketitle

\section{Introduction}
Knowledge about the mass composition of the Ultra High Energy Cosmic Rays (UHECRs) is essential when trying to 
explain the origin and acceleration mechanisms of the most energetic particles in the Universe. Due to their deflection in the 
galactic and extragalactic magnetic fields, a precise measurement of the mass (charge) of the incoming particles is absolutely 
necessary to correlate their arrival directions with the possible sources from the sky, in particular for light elements.  
After entering the Earth's atmosphere, the primary particles interact and produce
a huge number of secondary particles. 
These particles further interact or decay, producing in this way the so called Extensive Air Showers (EAS).
The electromagnetic longitudinal profile of the shower (the density of the charged particles as a function of the 
atmospheric depth) can be indirectly reconstructed using Fluorescence Detectors (FD) \cite{Abraham:2009pm}, by collecting the UV light emitted 
after the excitation of the nitrogen molecules by the secondary charged particles from the EAS. The atmospheric depth where the longitudinal profile 
reaches its maximum, $X_{\rm max}$, is an important observable sensitive to the mass of the primary particle
($X_{max}$ is related to ln $A$), the difference in $\langle X_{\rm max}\rangle$ between proton and iron induced showers at 
the same energy being $\sim 100$ g/cm$^{2}$  \cite{PhysRevD.90.122005}. 
This is an accurate method for reconstructing the longitudinal profile of EAS, 
and therefore of the $X_{\rm max}$ observable, but has the shortcoming of the low duty cycle of the FD 
(up to $15 \%$ \cite{ThePierreAuger:2015rma}). This is due to the fact that the UV light can be measured only in moonless nights 
in good atmospheric conditions.

The properties of the EAS were investigated using various techniques in different ground based experiments like the
Pierre Auger Observatory \cite{ThePierreAuger:2015rma}, Telescope Array \cite{AbuZayyad201287}, 
KASCADE \cite{Antoni:2005wq}, KASCADE-Grande \cite{Apel:2010zz}, HiRES \cite{Sokolsky201174}, 
AGASA \cite{Chiba:1991nf}, Yakutsk \cite{Knurenko:2006jt}. Despite huge efforts in the last decades, there are still large 
uncertainties in the reconstruction of the mass composition of UHECRs \cite{Aab:2017cgk, Aab:2016, Abbasi:2018nun}. 
First of all, this is due to the large uncertainties of the cross sections at these extreme energies ($E>$ \mbox{$10^{18}$ eV}), 
evaluated by extrapolation 
from lower energies accessible at LHC, in combination with the very low flux of 
UHECRs ($\sim 1$ particle km$^{-2}$ yr$^{-1}$ at $E = 10^{19}$ eV). 

Different complementary methods were developed \cite{Rebel:1994ed,Haeusler,Aab:2016enk,Aab:2014pza,Cazon:2004zx,Cazon:2003ar}
to obtain information about the primary particle mass, by making use of the signal recorded in the 
Surface Detectors (SD) \cite{AbuZayyad201287, Abraham:2010zz}. The duty cycle of such detectors is usually $\sim 100 \%$, thus, the statistics of the reconstructed 
events would increase considerably with respect to FD results.

One of these methods consists in the reconstruction of the longitudinal profile of muon production depth (MPD) from EAS, as proposed in 
\cite{Cazon:2004zx,Cazon:2003ar} in the context of the Pierre Auger Observatory. Using the arrival times $t_{\mu}$ of each muon in SDs relative 
to the time $t_{c}$ when the shower core reaches the ground, the heights on the shower axis where the muons were produced can be evaluated. 
Then, expressing the distribution of the production heights of all muons in units of atmospheric depth, the MPD longitudinal profile 
of the shower is obtained. The atmospheric depth where the maximum production rate of muons occurs, $X_{max}^{\mu}$, proved to be sensitive to the primary mass, the values of this observable for iron induced showers being lower than the values for proton induced showers at the same energy. The comparison of the experimental values of $X_{\rm max}^{\mu}$ with values obtained from simulation can also provide information useful to constrain the hadronic interaction models at highest energies \cite{Aab:2014dua}. For example, the experimental values of $\langle X_{\rm max}^{\mu}\rangle $ measured at the Pierre Auger Observatory indicate a disagreement between data and simulations based on the EPOS hadronic interaction model at high energies \cite{Aab:2014dua, PhysRevD.92.019903}. 
It should be mentioned that in the same energy range, measurements of the mass composition based on $\langle X_{\rm max}\rangle $ are in good agreement with MC simulations for 
two hadronic interaction models (QGSJetII-04 \cite{Ostapchenko:2011nk} and EPOS-LHC).

Another parameter sensitive to the primary mass is the total number of muons at ground level $N_{\text t}^{\mu}$ 
\cite{Haungs:2008zzb, Muller:2018zoc}. Its overall dependence on the primary mass and energy is given by the 
Matthews-Heitler model \cite{Matthews:2005sd}.

There is a special interest in the study of very inclined showers, because in this case the electromagnetic component is much attenuated due to the long atmospheric path, 
thus the particle density is dominated by the muonic component. The cleaner muonic component facilitates shower 
analysis. Also, including highly inclined showers increases the exposure of a particular experiment. Therefore, the problem of muon density in inclined showers was discussed in many papers, sometimes with the focus on specific features. Thus, 
an analytical description of the muon density was proposed \cite{Ave:2001xn} and the composition sensitivity, including the possibility of discriminating 
photon induced showers from particle induced showers was analyzed \cite{Ave:2002ve}. The shape of the distribution, governed by the 
geomagnetic field and the muon production depth distribution, was proposed as a method for mass discrimination \cite{Billoir:2015cua}.
Phenomenological parameterizations fitted to detailed Monte Carlo simulations were also proposed \cite{Ave:2000xs, Dembinski:2009jc}, and used for 
shower reconstruction \cite{Aab:2014gua}. Recently, a refined analysis of the shower development, including a new electromagnetic 
component due to low energy hadrons, complemented by a detailed detector simulation, resulted in a generalized description of the 
signal size in extensive air shower detectors \cite{Ave:2017uiv}.

Because the intrinsic shower properties are better described in the shower plane (also called normal plane), a procedure for mapping the coordinates of the 
muons observed in the detection plane onto the shower plane should be applied. The problem of mapping the particle densities 
onto the shower plane is of a complex nature. The purpose of this paper is restricted to the study of the effect of the procedure of mapping the muon density from the ground 
plane onto the shower plane on the reconstruction of the muon production depth and of the number of muons contributing to the MPD profile. 

We will reconstruct the MPD profiles using muon produced on the shower core and strike the ground plane, whose corresponding coordinates into the normal plane lie in specific radial ranges. These corresponding coordinates are obtained following two approaches. In the first one we use a simple orthogonal projection from the ground plane to the normal plane, without including the differential attenuation effect. We will call this method (ort). The second projection method follows the true incoming directions of muons and takes into account the differential attenuation effect from Eq. \ref{correct_density}. We call this method (att).

In this exploratory study, based on CORSIKA simulations \cite{corsika,corsika1}, we consider only the muonic component of the shower, which would be relevant in the case of shielded detectors or for inclined showers far from 
the core, after correcting for the electromagnetic background. The zenith angles are restricted to the range $\theta = [37^{\circ} - 60^{\circ}]$, 
the high limit being chosen to avoid the effect of geomagnetic deflection.

In Section \ref{rest_azim} we describe the method applied for mapping the particle density from the ground plane to the shower plane 
which restores the azimuthal symmetry of the distribution of the muons around the shower axis. 
In Section \ref{eval_X_N} we evaluate the $X_{\rm max}^{\mu}$ and the number $N_{\mu}$ of muons which contributed to the constructed MPD. These quantities are computed
using two projection methods onto the normal plane 
(orthogonal projection and projection along the particle momentum). 
In Section \ref{results} we present the implication of the projection methods on the estimation of $X_{\rm max}^{\mu}$ 
and $N_{\mu}$. Section \ref{conclusions} concludes the paper.

\section{Restoring the azimuthal symmetry around the shower axis} \label{rest_azim}

\begin{figure}
\centering
\includegraphics[height=.3\textheight]{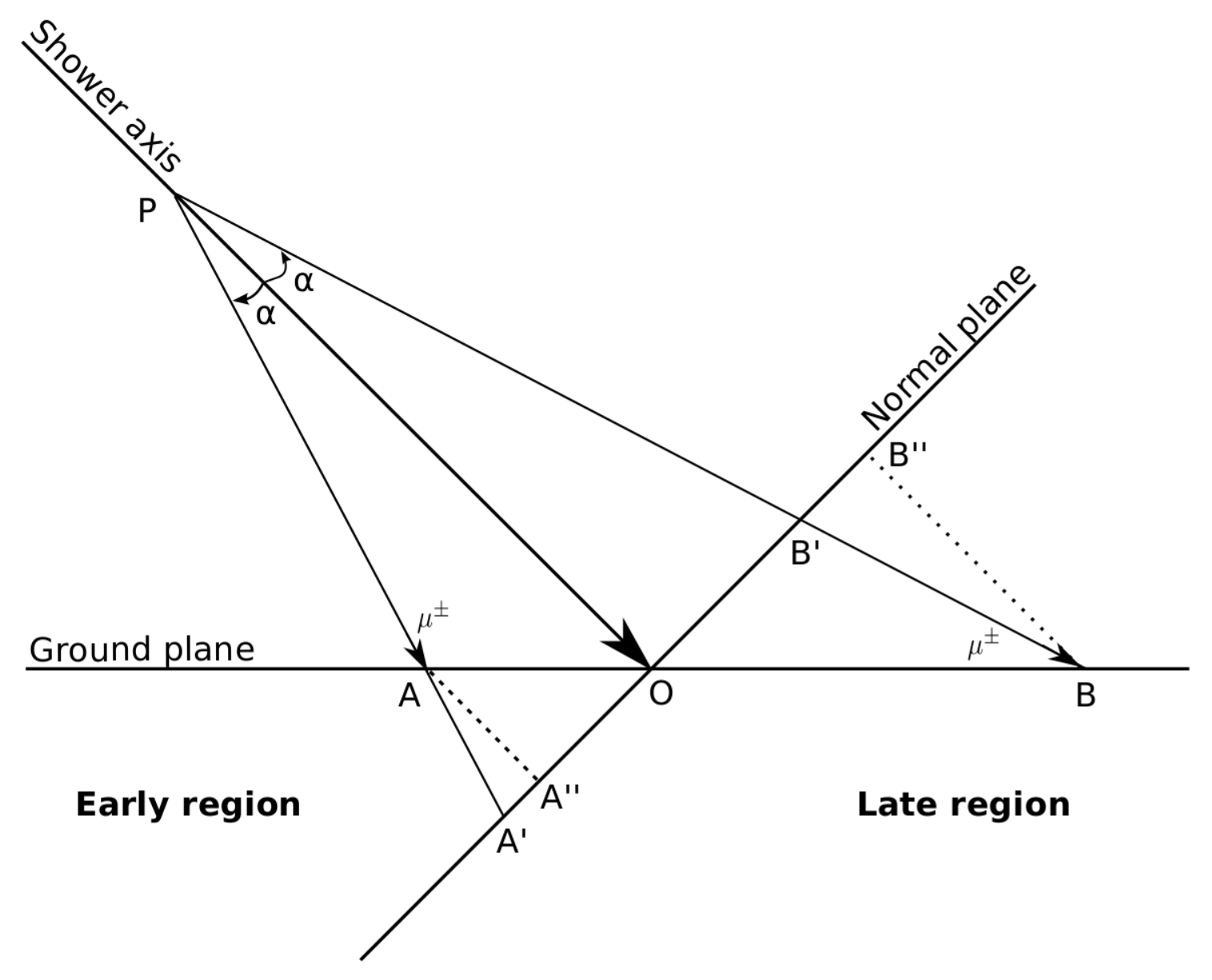}
\caption{Schematic view of the shower geometry. Particles produced below the shower axis hit the ground plane in the $early \ region$, 
while those produced above will arrive in the $late \ region$. We emphasize the two projection methods: the simple orthogonal projection 
from the ground plane onto the normal plane (dashed lines $A \rightarrow A''$ and $B \rightarrow B''$), and the projection along the 
incoming directions (continuous lines $A \rightarrow A'$ and $B \rightarrow B'$).}
\label{projection_}
\end{figure}

$X_{\rm max}^{\mu}$ is evaluated experimentally 
by taking into account the muons which contribute to the MPD and whose coordinates in the plane perpendicular to the shower axis 
lie in a specific radial range. On the ground plane, the azimuthal distribution of muons is asymmetric for inclined showers \cite{Ave:2001xn} - \cite{Ave:2017uiv}. 
This asymmetry arises from the geometry of the shower axis, attenuation effects and deflections in the geomagnetic field.
In the case of inclined showers with $\theta = [37^{\circ} - 60^{\circ}]$, the length of the muons' trajectory through the geomagnetic 
field is not sufficient for introducing significant asymmetry, therefore our analysis is focused on geometric and attenuation effects.
The particles produced below the shower axis strike the ground first, in the $early \ region$, while those produced above the shower axis, in the $late \ region$,
will experience additional attenuations or decay. 
Suppose that two muons are produced in the
point $P$ on the shower core (see Fig. \ref{projection_}), at the same angle $\alpha$ relative to the shower axis, and hit the ground plane in the points 
$A$ and $B$ respectively. To analyze the muons from a specific radial range in the normal plane, the coordinates from the 
ground plane should be mapped onto the normal plane. After a simple orthogonal projection (dashed lines in Fig. \ref{projection_}),
the point $A$ translates into $A''$ and point $B$ into $B''$. One can observe that $OB''$ is larger than $OA''$, which 
means that using this projection method, the density of muons as a function of the distance to the shower axis is distorted. 
The consequence of this distortion may be a wrong estimation of the number of muons produced in the point $P$ if, for example, a too small radial range is considered.
By projecting the particles along their momentum vector ($A \rightarrow A'$ and $B \rightarrow B'$) and including also the attenuation 
probability, the azimuthal symmetry is restored at least approximately.
Therefore in this work the density in the normal plane is obtained by projecting along the muon momentum, including 
also differential attenuation effects \cite{Sima:2011zz}.

From now on, in this study we will refer only to the secondary muons from the EAS.
The details regarding the evaluation of the attenuation probability will be given in Section \ref{lambda}.

\subsection{Simulations}

We analyze a set of CORSIKA simulations comprising showers induced by protons and iron nuclei as primary particles, including EPOS as the hadronic interaction 
model at high energies \cite{Pierog:2009zt} and FLUKA at low energies \cite{Ferrari:898301}. 

The simulations were done for $p$ and $Fe$ primary particles, at two fixed energies, $E=10^{19}$ and $10^{20}$ eV, at $4$ zenith 
angles $\theta = 37^{\circ}, 48^{\circ}, 55^{\circ}$ and $60^{\circ}$ in the conditions of the Pierre Auger Observatory (geomagnetic 
field and height of the observation level).
In each case $120$ showers were prepared, with random azimuth 
angles $\phi$, thus in total $1920$ showers were analyzed.
In order to reduce computation time and the output size, the thinning algorithm \cite{Hillas:1997tf} is implemented in the CORSIKA simulation code by replacing, in certain conditions, a bunch of $n$ secondary particles of the same type, with a single particle with a weight $n$. The thinning level is defined as $\epsilon_{th} = E / E_{0}$, where $E_{0}$ is the energy of the primary particle which initiated the shower and $E$ represents the summed energy of the secondary particles exposed to the thinning algorithm.
The required condition for activating the thinning algorithm is $E < \epsilon_{th} E_{0}$. In our simulations we used $\epsilon_{th} = 10^{-6}$.
The same set of showers has been used in our previous study of the muon production depth \cite{Arsene:2016hoo}.
Experimental uncertainties were not included in the analysis. 

\subsection{Evaluation of the attenuation parameter $\lambda$}
\label{lambda}

As proposed in \cite{Sima:2011zz}, to restore the azimuthal symmetry of the lateral distribution of the secondary particles around the shower axis in the 
normal plane, the probability of particles to survive from the ground plane to the normal plane in the $early \ region$ 
(from $A \rightarrow A'$), but also the attenuation from $B'$ to $B$ in the $late \ region$ (see Fig. \ref{projection_}) should be accounted for. Consequently, the procedure for estimating the lateral density of muons 
in the shower plane is the following. For a muon arriving at point $\vec{r}$ in the ground plane, the point $\vec{r}\ '$ of incidence of its trajectory on the normal plane is determined. A weight factor $w$, 

\begin{equation}\label{correct_density}
w = e^{ \ -s \lambda D }
\end{equation}
is assigned to this muon. In the equation $D=|\vec{r}\ ' -\vec{r}|$ is the distance along the muon trajectory between the two planes, $\lambda$ is the differential attenuation parameter, and $s$ distinguishes between the $late \  region$ ($s=-1$) and the $early \ region$ ($s=1$). The density in the normal plane is evaluated by summing the weight factors of the muons with $\vec{r}\ '$ in a given domain and dividing by the area of that domain.

The role of the Eq. \ref{correct_density} is to ensure that muons which hit the ground level at point $A$ (\textit{early region}) and don't have enough energy to propagate to point $A^{'}$, or are expected to decay, to be weighted by the factor $w$ which takes into account the differential attenuation and decay. In this situation the weight factor for these muons is smaller than 1. It means that around point $A^{'}$ the muon density will be smaller than the density around point $A$. In the same way we proceed for the \textit{late region}. Note that in this case the weight factors are greater than 1, which translates in a higher density around point $B^{'}$ in comparison with point $B$.

The value of $\lambda$ depends on the zenith angle of the primary cosmic ray, the muon momentum, 
and the muon position on the ground. The individual muon momentum cannot be obtained experimentally, thus we have to impose 
a threshold value for the muon's momentum $p_{\rm th}$, and to consider in the analysis muons with $p > p_{\rm th}$. For example, the muon 
detection threshold in the water-Cherenkov tanks \cite{Ave:2007zz} from the Pierre Auger Observatory is $\sim 0.3$ GeV/c 
(the threshold of the Cherenkov effect for muons in water), whereas after the upgrade with the scintillator detectors \cite{Aab:2016vlz}, different values
for $p_{\rm th}$ could be considered. 
In this context, it should be mentioned that a simulation study regarding the impact of the $p_{\rm th}$ on the sensitivity of $(N_{\rm t}^{\mu},X_{\rm max})$ observables
to the primary mass was recently done \cite{Muller:2018zoc}.

We evaluate the parameter $\lambda (A, \theta, p_{\rm th}, R)$ ($A$ stands for primary $p$ or $Fe$) for each shower as follows:
\begin{itemize}
\item[-]
each muon is projected from the ground onto the normal plane along its incoming direction;
\item[-]
the muons from different radial ranges $R = [500 - 1000], [1000 - 1500], ..., [3500 - 4000]$ m in the normal plane are separately analyzed;
\item[-]
in each radial range, several muon momentum thresholds $p_{\rm th} = 0.3,\ 0.8,\ 1.3,\ 1.8$ and $2.3$ GeV/c are set;
\item[-]
each muon is weighted according to Eq. \ref{correct_density}, while $\lambda$ is varied in a specific range until the azimuthal 
distribution of muons becomes symmetric; this is done by performing a linear fit of the azimuthal distribution following a $\chi^{2}$ 
minimization;  
\item[-] finally, the set of $\lambda (A, \theta, p_{\rm th}, R)$ values which best restore the azimuthal symmetry is obtained.
\end{itemize}

\begin{widetext}

\begin{figure}
 \centering
\includegraphics[height=.5\textheight]{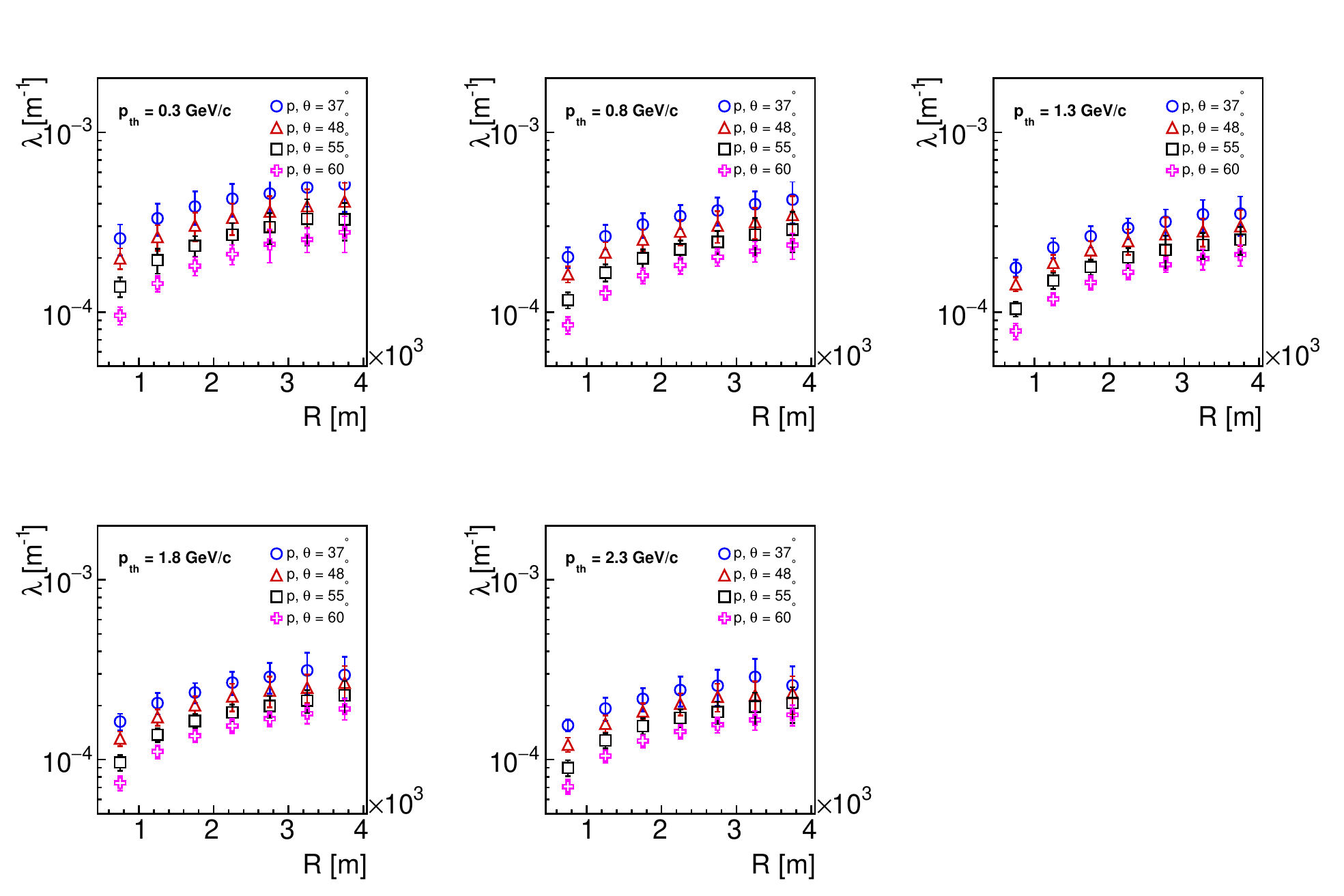}
\caption{Evolution of $\lambda$ as a function of the distance to the shower axis in the normal plane for $p$ induced showers 
at \mbox{$E =10^{20}$ eV}, 4 zenith angles $\theta = 37^{\circ}, 48^{\circ}, 55^{\circ}$ and $60^{\circ}$, and 5 muon momentum 
thresholds $p_{\rm th} = 0.3,\ 0.8,\ 1.3,\ 1.8$ and $2.3$ GeV/c. Each point corresponds to a radial bin $\Delta R = 500$ m.}
\label{lambda_p_e_20_}
\end{figure}

\end{widetext}

The attenuation parameter $\lambda (p, \theta, p_{\rm th}, R)$ as a function of 
the distance to the shower axis in the normal plane for $p$ induced showers at $E = 10^{20}$ eV, for different zenith angles and 
different values of $p_{\rm th}$ is represented in Fig. \ref{lambda_p_e_20_}.
As can be seen, the $\lambda$ parameter depends on the muon momentum threshold $p_{\rm th}$ and zenith 
angle of the shower axis $\theta$, having larger values for lower values of $p_{\rm th}$ and $\theta$. 
The results obtained for $p$ showers at $10^{19}$ eV, as well as for $Fe$ induced showers at both primary energies $10^{19}$ and $10^{20}$ eV look
similar to those presented in Fig. \ref{lambda_p_e_20_}. The main difference is given by the larger uncertainties of the $\lambda$ 
parameter for $p$ showers, due to the larger shower-to-shower fluctuations.

\begin{figure}
\centering
\includegraphics[height=.30\textheight]{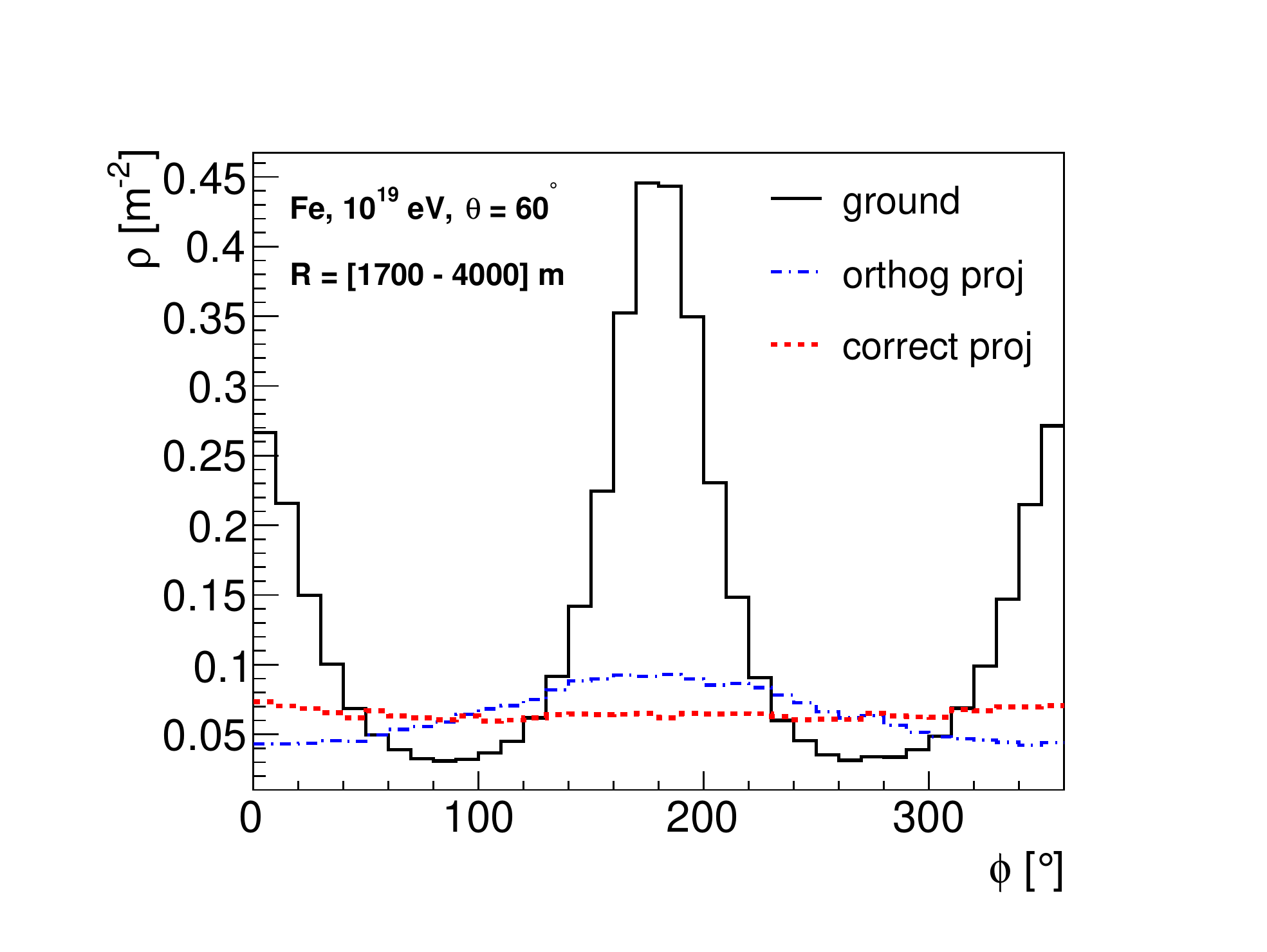}
\caption{The distribution of the muon density from an $Fe$ induced shower with $E = 10^{19}$ eV and zenith angle $\theta = 60^{\circ}$, simulated with 
CORSIKA. In the ground plane (continuous line) the distribution is strongly asymmetric due to the shower geometry. A simple 
orthogonal projection does not restore the symmetry (blue dash-dot line). After the projection along the muons momentum and including the attenuation 
correction, the azimuthal symmetry of the density is restored (red dashed line).  Radial range in the shower plane: $R = [1700 - 4000]$ m.}
\label{asimmetry_}
\end{figure}

After the estimation of $\lambda (A, \theta, p_{\rm th}, R)$, 
the muon density 
in the normal plane is obtained by projecting along the muons' incoming directions and applying the weight factor given in Eq. \ref{correct_density} to correct for differential attenuation.
In Fig. \ref{asimmetry_}, the distribution 
of the muons density for an $Fe$ induced shower at $\theta = 60^{\circ}$, taking into account only the muons from the radial range $R = [1700 - 4000]$ m is displayed. The coordinate system in the normal plane is defined as follows: the intersection of the horizontal plane with the normal plane defines the Y axis, while the X axis is defined by the intersection of the vertical plane containing the shower axis with the normal plane. The positive direction of the X axis corresponds to forward directions in the horizontal plane. The $\phi$ angle is defined as usually in the normal trigonometrical sense in the shower plane.

We represent the muon density around the shower axis in the ground plane (asymmetric due to the shower geometry),
in the normal plane after an orthogonal projection and after the projection along the muons momentum, with the correction for differential attenuation 
included. As can be seen, the azimuthal distribution of muons in the normal plane obtained using the latter procedure is symmetric, whereas the orthogonal projection 
maintains an asymmetry of the distribution with an amplitude of about $\Delta \rho \simeq 35 \%$ .

\section{Evaluation of $\mathbf{X_{\rm max}^{\mu}}$ and $\mathbf{N_{\mu}}$} \label{eval_X_N}

The MPD longitudinal profile of EAS can be reconstructed experimentally on the basis of the signal induced in the SDs; the advantage is the high 
duty cycle $\sim 100 \%$. The method is suitable for experiments which can record the arrival times of the secondary particles in SDs, 
for example the Pierre Auger Observatory. The muons produced on the shower core, propagate almost in a straight line through the 
atmosphere, suffering negligible deviations in the geomagnetic field (zenith angle $\theta < 60^{\circ}$). 
We apply the procedure described in \cite{Arsene:2016hoo} to calculate the muon 
production height in units of [g cm$^{-2}$] for individual muons. 
Only the muons produced on the shower core 
(i.e., for which the difference in the arrival time of muons at ground $t_{\mu}$ relative to the time when the shower core reaches the ground $t_{c}$, is consistent with the difference in their pathlengths \cite{Arsene:2016hoo}), and which 
have the radial coordinates in the normal plane belonging to  specific radial ranges, are considered. The distribution of the MPD is constructed by applying to the production depth of 
each muon the corresponding weight factor. The resulting MPD profile is fitted with the Gaisser-Hillas function \cite{GH} and its maximum,
$X_{\rm max}^{\mu}$, is evaluated. A quality criterion is additionally imposed in the determination of  $X_{\rm max}^{\mu}$, namely only 
the showers whose fit quality satisfy $\chi^{2}/ndf < 2.7$ are included in the analysis. This cut had been adopted from 
previous studies \cite{Arsene:2016hoo}, but it is worth mentioning that varying the cut level has a negligible effect on the $X_{\rm max}^{\mu}$ estimates,  affecting slightly only their uncertainties.

We evaluated also $N_{\mu}$, which represents the number of muons which contribute to the MPD profile (not the total number of muons, $N_{\text t}^{\mu}$, from the EAS). This parameter 
depends on the inclination angle of the shower axis and the radial range of interest in the normal plane. Both parameters, $X_{\rm max}^{\mu}$ and $N_{\mu}$, depend also 
on the muon momentum threshold imposed when reconstructing the MPD profile. 

In the next section we present the main differences between the values of $X_{\rm max}^{\mu}$ and $N_{\mu}$ obtained using the two 
projection methods (orthogonal projection and the method described in Section \ref{rest_azim}).

\section{Results}\label{results}

The $X_{\rm max}^{\mu}$ and $N_{\mu}$ parameters were evaluated for each simulated shower, considering different values for the muon momentum threshold 
$p_{\rm th} = 0.3,\ 0.8,\ 1.3,\ 1.8$ and $2.3$ GeV/c, and taking into account only the muons whose coordinates in the normal plane lie in 
three radial ranges $R = [1300 - 4000]$ m, $R = [1500 - 4000]$ m and $R = [1700 - 4000]$ m. 
For each analyzed shower we obtain two sets of ($X_{\rm max}^{\mu}$, $N_{\mu}$) parameters corresponding to the different projection methods: 
orthogonal projection of muons from the ground plane to the normal plane ($X_{\rm max}^{\mu}(\rm ort)$, $N_{\mu}^{\rm ort}$), and projection along 
their incoming direction, corrected for differential attenuation effects ($X_{\rm max}^{\mu}(\rm att)$, $N_{\mu}^{\rm att}$).

In order to quantify the effect of the projection methods on the two observables $X_{\rm max}^{\mu}$ and $N_{\mu}$, we define the quantities 
$\Delta X_{\rm max}^{\mu} = X_{\rm max}^{\mu}(\rm att) - X_{\rm max}^{\mu}(\rm ort)$ and the ratio $N_{\mu}^{\rm att} / N_{\mu}^{\rm ort}$.

\begin{widetext}

\begin{figure}
 \centering
\includegraphics[height=.55\textheight]{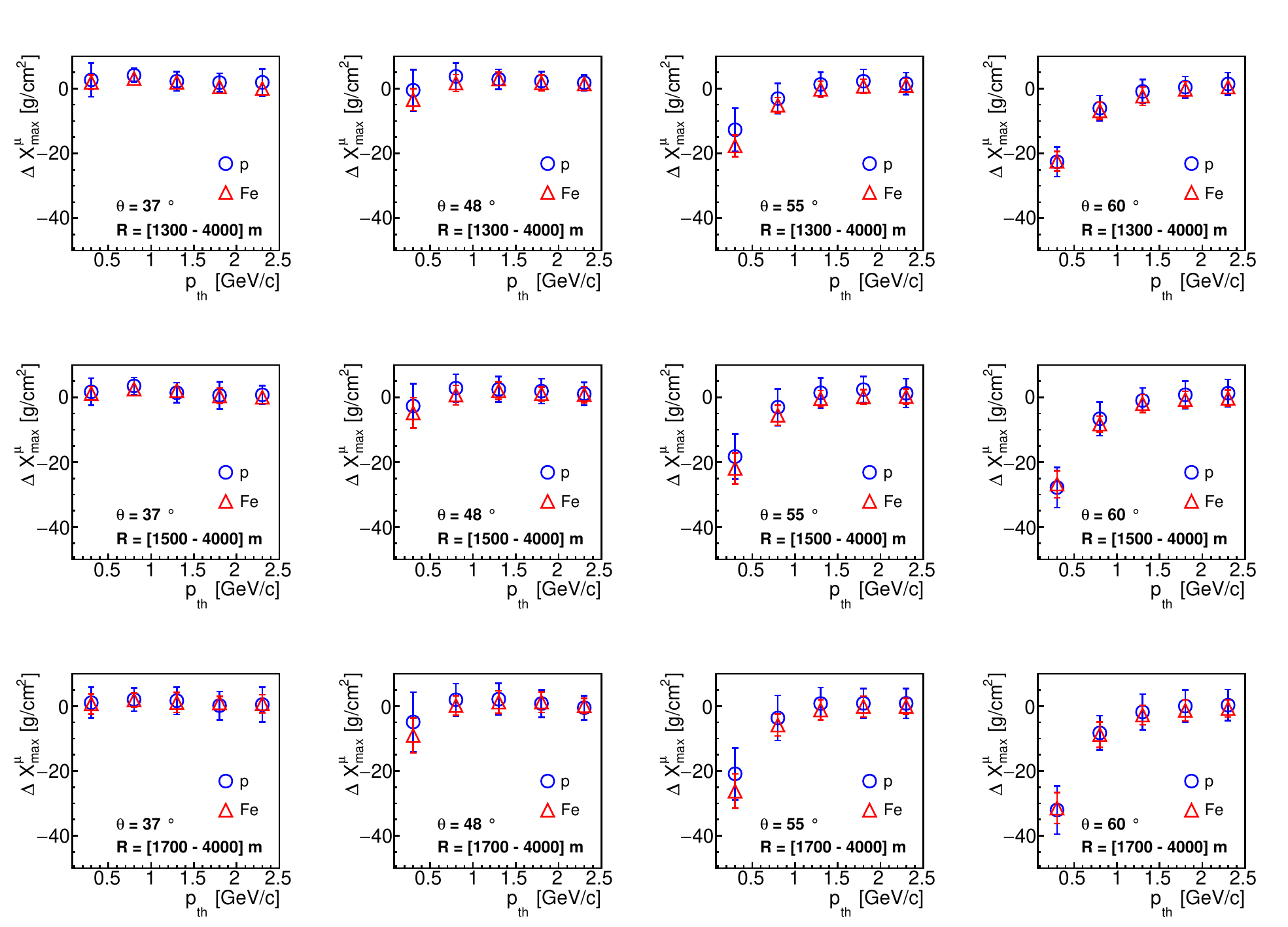}
\caption{$\Delta X_{\rm max}^{\mu}$ as a function of $p_{\rm th}$ for $p$ and $Fe$ induced showers at $E = 10^{20}$ eV and different zenith 
angles $\theta = 37^{\circ}, 48^{\circ}, 55^{\circ}$ and $60^{\circ}$. 
The radial ranges in the normal plane of the muons analyzed are $R = [1300 - 4000]$ m \textit{(top)}, $R = [1500 - 4000]$ m 
\textit{(center)} and $R = [1700 - 4000]$ m \textit{(bottom)}.
The error bars represent the uncertainty given by the shower-to-shower fluctuations.}
\label{delta_pth_e20_}
\end{figure}

\end{widetext}

In Fig. \ref{delta_pth_e20_} we plot the dependence of $\Delta X_{\rm max}^{\mu}$ on $p_{\rm th}$ for showers with 
$E = 10^{20}$ eV and different zenith angles for the muons belonging to the three radial ranges in the normal plane.
It can be seen that the reconstruction of $X_{\rm max}^{\mu}$ is strongly dependent on the chosen projection method.  
We found smaller $X_{\rm max}^{\mu}$ values when the attenuation effect is considered in the analysis.
The difference $\Delta X_{\rm max}^{\mu}$ is larger for small muon momentum thresholds, large zenith angles and higher radial ranges.
The largest difference $\Delta X_{\rm max}^{\mu} \simeq - 32$ g/cm$^2$, representing almost  $50 \%$ of the proton-iron separation, is obtained for $p_{\rm th} = 0.3$ GeV/c, $\theta = 60^{\circ}$ and $R = [1700 - 4000]$ m.
This value should be compared with the experimental resolution (systematic uncertainty) of $X_{\rm max}^{\mu}$ measurement at the Pierre Auger Observatory ($\sim 17$ g/cm$^{2}$) \cite{Collica:2016bck}.
For larger $p_{\rm th}$ values, the difference $\Delta X_{\rm max}^{\mu}$ becomes negligible, but setting a high value of $p_{\rm th}$ in data analysis 
implies having a poor 
statistics in the MPD distributions, which translates in large uncertainties for the estimation of $X_{\rm max}^{\mu}$. 

What changes the behavior of the reconstructed MPD profiles is the way (i.e. (ort) or (att) projection method) we select the muons which contribute to the MPD, whose coordinates into the normal plane lie in a specific radial range, and on the other hand, the corrections for the differential attenuation effects. 
The lateral distribution in the normal or in the ground plane of the muons produced
in a given atmospheric depth, depends on the production depth. That means that the
reconstructed production depth depends on the radial range in which the muons used for
reconstruction are sampled. The projection method proposed in this paper, (att), aims to select a more homogeneous sample of muons to be used for production depth reconstruction. Indeed we consider that the sample of muons in a given radial range obtained by the proposed projection method is more
homogeneous than the sample of muons in the same radial range obtained by orthogonal
projection, because in the latter case the dependence on the particular azimuthal coordinates is strong. Of course the distribution of production depths obtained using our method
is not identical with the ideal distribution of the production depths which could be obtained
by analyzing all the muons, but if it is sensitive to the nature of the primary particle it could
be useful in shower reconstruction.

Concerning these results, we emphasize the following issues. In order to extract information on mass composition from the experimental values 
of $X_{\rm max}^{\mu}$, the experimental values should be compared with corresponding values obtained from simulations. In this approach, 
besides the mapping procedure, several other reconstruction steps,
which may act somewhat differently on experimental than on simulation data 
\cite{Aab:2014dua, PhysRevD.92.019903}, are involved. As a consequence, the 
distribution of the values of $X_{\rm max}^{\mu}$ becomes broader, due to the effect of several factors, like e.g. the unthinning algorithm 
applied to the simulation data. In order to get a meaningful evaluation of the effect of the mapping procedure, the results presented 
in this work were obtained by applying the two mapping procedures to all the muons simulated by CORSIKA which have the radial coordinate 
in the shower plane in the given radial ranges. Thus, the values of $X_{\rm max}^{\mu}$ reported here, which are based on the 
contributions to the MPD of all the muons, not only of the muons which happen to hit the detectors, have lower uncertainties. Due to this fact 
we consider that the values of $X_{\rm max}^{\mu}(\rm att)$ and $X_{\rm max}^{\mu}(\rm ort)$ obtained in this way represent a better 
reference for inferring the mass composition from the experimental values of $X_{\rm max}^{\mu}$.
We already checked that by analyzing the same sets of CORSIKA simulations taking into account only muons which hit an array of detectors similar to those from the Pierre Auger Observatory, the statistical spread of the results is higher. In view of the narrower shower to shower fluctuations of the $X^{\mu}_{max}$ values obtained by simulations including all the muons, we consider justified to propose these $X^{\mu}_{max}$ values as reference in analyzing the experimental results.

The restoration of the azimuthal symmetry
of the density in the shower plane is clearly demonstrated and this feature is very useful
in every respect in the shower reconstruction. This is especially so when analyzing data
pertaining to few locations around the shower core, obtained from detectors with specific
radial and angular coordinates; we expect a higher bias of the reconstructed quantities if the
information comes from observed densities strongly dependent on the azimuthal coordinates
than in the case when the azimuthal dependence is removed.

If the mass composition would be inferred 
from the same set of $X_{\rm max}^{\mu}$ values obtained from experiment by comparison with the values $X_{\rm max}^{\mu}(\rm att)$ 
instead of $X_{\rm max}^{\mu}(\rm ort)$, the lower values of $X_{\rm max}^{\mu}(\rm att)$ would suggest a lighter primary composition 
in comparison with the values obtained by reference to $X_{\rm max}^{\mu}(\rm ort)$. In this context, we mention that the results presented 
in reference \cite{Aab:2014dua, PhysRevD.92.019903} were obtained using the orthogonal projection.
We do not know for sure whether the application of the proposed mapping procedure to experimental data will shift the $X^{\mu}_{max}$ values to exactly the same extent as in the case of the simulations; we suspect that this is not the case. But the conclusion that the comparison of a given set of $X^{\mu}_{max}$ values extracted from experimental data with the lower $X^{\mu}_{max}(att)$ values obtained from simulation would suggest a lighter mass composition than in the case when the orthogonal projection would be used for obtaining $X^{\mu}_{max}$ from simulations, is correct.

\begin{widetext}

\begin{figure}
 \centering
\includegraphics[height=.55\textheight]{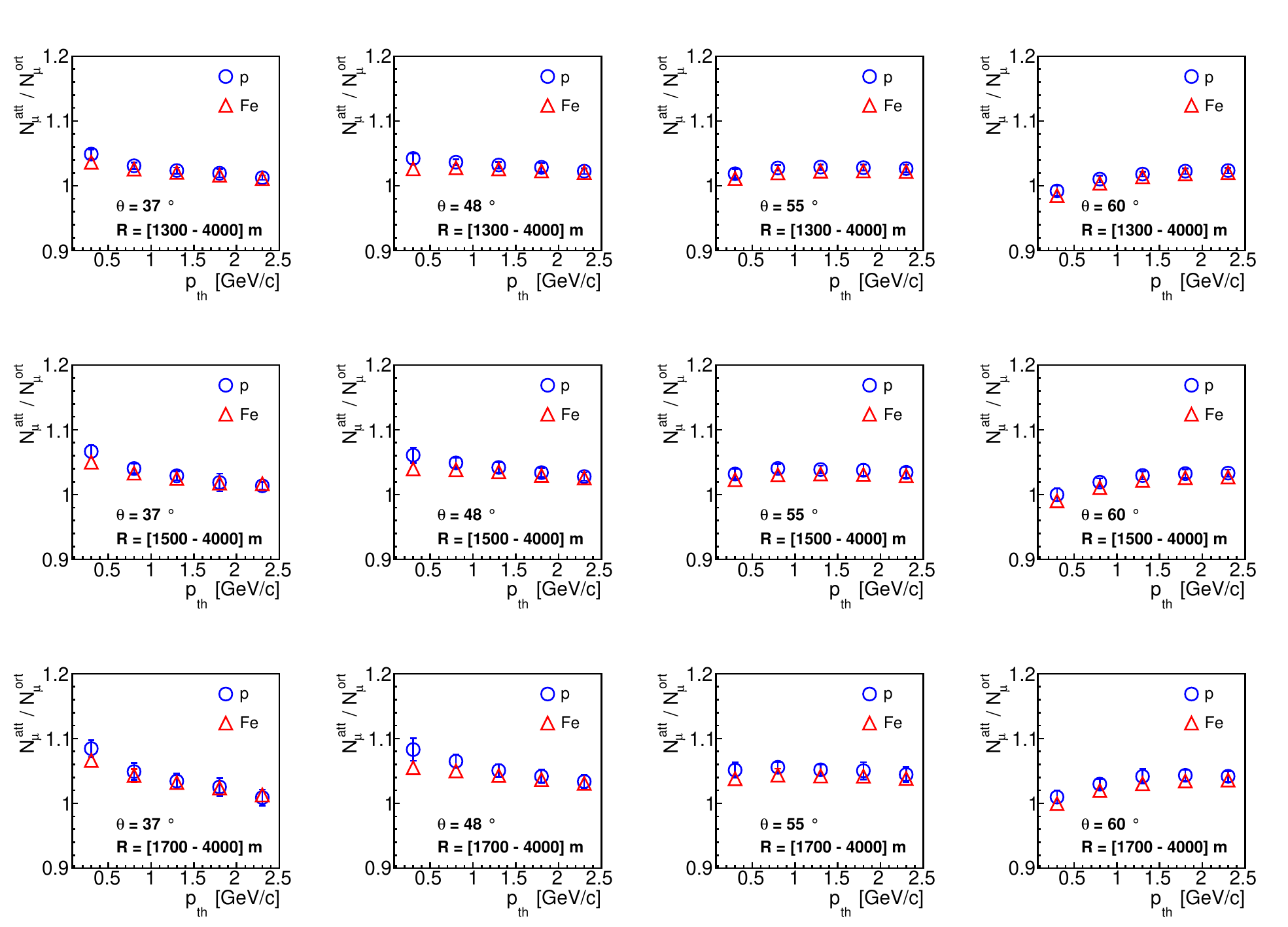}
\caption{$N_{\mu}^{\rm att} / N_{\mu}^{\rm ort}$ as a function of $p_{\rm th}$ for $p$ and $Fe$ induced showers at $E = 10^{20}$ eV and different zenith 
angles $\theta = 37^{\circ}, 48^{\circ}, 55^{\circ}$ and $60^{\circ}$. Muon radial ranges in the 
normal plane: $R = [1300 - 4000]$ m \textit{(top)}, $R = [1500 - 4000]$ m \textit{(center)} and $R = [1700 - 4000]$ m \textit{(bottom)}.
The error bars represent the uncertainty given by the shower-to-shower fluctuations.}
\label{ratio_Nmu_e20_}
\end{figure}

\end{widetext}

We also investigated the influence of the two projection methods on the number of muons contributing to the MPD profiles.
In Fig. \ref{ratio_Nmu_e20_} we plot the dependence of $N_{\mu}^{\rm att} / N_{\mu}^{\rm ort}$ as a function of $p_{\rm th}$, for the same showers 
analyzed in Fig. \ref{delta_pth_e20_}. The biggest difference is observed for showers with smaller zenith angles and smaller values 
of $p_{\rm th}$. The number of muons  $N_{\mu}^{\rm att}$ is larger than $N_{\mu}^{\rm ort}$ with an amount of $\sim 10 \%$ for showers induced 
at $\theta = 37^{\circ}$ and $R = [1700 - 4000]$ m.

The results obtained for showers with primary energy $E = 10^{19}$ eV are quite similar to those obtained for $E = 10^{20}$ eV, for 
both observables $\Delta X_{\rm max}^{\mu}$ and $N_{\mu}^{\rm att} / N_{\mu}^{\rm ort}$.

\section{Summary and Conclusions} \label{conclusions}

In this study a procedure for the reconstruction of the distribution of the MPD is presented. The study is based on CORSIKA simulations at two fixed energies and four angles of incidence. Production depth of individual muons is evaluated on the basis of arrival times; the muon incoming direction is also obtained. The lateral muon density in the plane normal to the shower axis is constructed by 
projecting the muon impact point in the ground plane along the muon momentum and applying a correction for differential attenuation, by assigning a weight factor associated with attenuation to each muon. This procedure restores the axial symmetry of the lateral density of muons in the normal plane. The distribution of the MPD is constructed using the weight factors of the muons from selected  radial ranges (typically $R = [1700 - 4000]$ m) in the plane normal to the shower axis. 

Further, the distributions of the MPD were fitted with Gaisser-Hillas functions and the values of $X_{\rm max}^{\mu}$ were estimated. These values, as well as the values of the number of muons $N_{\mu}$ obtained by integrating the lateral density of muons which contributed to the MPD 
from a given radial range in the shower plane, were compared with the corresponding values obtained by applying the orthogonal projection. We found smaller $X_{\rm max}^{\mu}$ values when the refined method including the attenuation corrections was applied. 
In conditions similar to those from the Pierre Auger Observatory 
($p_{\rm th} = 0.3$ GeV/c and $R = [1700 - 4000]$ m) we found $\Delta X_{\rm max}^{\mu} \simeq - 32$ g/cm$^2$ which represents almost $50 \%$ 
of the iron-proton separation.

The lower values of the $X_{\rm max}^{\mu}$ in the simulations when compared with given experimental values of  $X_{\rm max}^{\mu}$ would be interpreted as a lighter mass composition of the cosmic rays.

Our findings suggest that the method of mapping the muon density from the ground plane to the shower plane has a significant effect on the MPD distributions constructed for muons belonging to specific radial bins in the shower plane. In order to avoid a biased estimation, the same mapping procedure, including corrections for attenuation effects, should be applied both to experimental data and to simulations when using the $X_{\rm max}^{\mu}$ observable for the determination of the mass composition of the cosmic rays.

Finally, we emphasize that in the present study no experimental details were considered.  In actual experiments 
larger uncertainties, related to the muon arrival time uncertainties, the primary energy reconstruction, and due to the poor statistics in the MPD distributions, are expected.
In this context, we would like to emphasize that the Auger Upgrade "AugerPrime" will include improved electronics, capable to achieve 
better timing accuracy and faster ADC sampling \cite{Martello:2017pch}. With improved values of the arrival times, the uncertainty 
of the MPD reconstruction will decrease and the effect described in this paper should be accounted for.


\subsection*{Acknowledgments}

We would like to thank our colleagues from the Pierre Auger Collaboration for many interesting and useful discussions. N. A. acknowledges financial support from the LAPLAS
VI program of the Romanian National Authority for Scientific
Research (CNCS-UEFISCDI). The work of O. S. was supported by a grant of the Romanian Ministery of Research and Innovation, CCCDI - UEFISCDI, project number PN-III-P1-1.2-PCCDI-2017-0839/19PCCDI/2018, within PNCDI III.

\bibliography{azimuthal_symmetry}

\end{document}